\documentclass[a4paper]{aa}
\usepackage{graphicx}
\usepackage{natbib}
\usepackage{psfig}
\usepackage{amsmath}
\usepackage{aas_macros}
\usepackage{longtable}
\bibpunct{(}{)}{;}{a}{}{,}

\def\v{~variability~}
\def\N{~NGC~2516~}

\usepackage[center]{subfigure}

\begin{document}
\title{X-ray variability of \N stars in the XMM-Newton observations}

\titlerunning{X-ray variability of \N stars}
\authorrunning{Marino et al.}

\author{A. Marino\inst{1}
\and
G. Micela\inst{2}
\and
I. Pillitteri\inst{1}
\and
G. Peres\inst{1}
}

\offprints{A. Marino  e-mail: marino@astropa.unipa.it}

\institute{DSFA, Universit\`a di Palermo, Piazza del Parlamento 1, 90134 Palermo - ITALY;
\and
INAF - Osservatorio Astronomico di Palermo,
Piazza del Parlamento 1, 90134 Palermo - ITALY;
}

\date{Received 9 December 2005 / accepted 01 May 2006 }

\abstract{We present the characteristics of the X-ray variability
of stars in the  cluster \N as derived from  XMM-Newton/EPIC/pn data. 
The  X-ray variations on short (hours), medium (months), and long (years) 
time scales have been explored.
We detected  303 distinct X-ray sources by analysing six EPIC/pn observations;
194 of them are members of the cluster. 
Stars of all spectral types, from the early-types to the late-M dwarfs, 
were detected.  The Kolmogorov-Smirnov test applied to the X-ray photon 
time series shows that, on short time scales, only a relatively small fraction
(ranging from 6\%  to 31\% for dG and dF, respectively) of the members of \N 
are variable with a confidence level $\geq$99\%; however, it is possible  that 
the fraction is small only because of the poor statistics.
The time X-ray amplitude distribution functions (XAD) of a set of dF7-dK2 stars, 
derived on short (hours) and medium (months) time scales, seem to suggest that
medium-term variations, if present, have a much smaller amplitude than those 
on short time scales; a similar result is also obtained  for dK3-dM stars.
The amplitude variations of late-type stars in \N are consistent with those 
of the coeval Pleiades stars.
Comparing these data with those of ROSAT/PSPC, collected 7-8 years earlier, 
and of ROSAT/HRI, just 4-5 years earlier, we find no evidence of significant  
\v on the related time scales, suggesting that long-term variations due to 
activity cycles similar to the solar cycle are not common among young stars.
Indications of spectral \v was found in one star whose spectra at three epochs
 were available.

}
 
\maketitle

\section{Introduction}
Observations using the {\em Einstein} observatory, nearly  30 years ago, 
showed the ubiquity of stellar X-ray emission throughout the HR diagram
\citep{Va81}.
X-ray emission from stars has been attributed to several physical
mechanisms depending on the mass of the star. Going from high to low mass, 
they include winds, 
dynamo, and turbulent dynamo.
Variability studies are one of the best ways to explore these mechanisms 
and to infer physical conditions of the regions where X-ray emission originates
and their evolution along stellar life  \citep[e.g.][]{Mont83, Amb87, Ster95, 
Mar03, Mar03a, Pi05}.
In this context, open clusters, providing large, chemically homogeneous, and 
precisely dated samples of stars, are ideal laboratories for studying coronal 
emitters and for constraining the X-ray-generating mechanisms. 
Comparative studies of the \v properties of homogeneous classes are excellent 
for diagnosing and  determining the origin of the X-ray emission.

In this perspective we have analysed the X-ray variability properties
of  homogeneous samples (with respect to mass and age): dF7-dK2 and  dK3-dM 
stars in the solar  neighbourhood, dF7-dK2, and dK3-dM of Pleiades observed 
with ROSAT-PSPC \citep{Mar00, Mar02, Mar03a}. These studies  show that the 
short-term \v of solar-type stars decreases with age (together with luminosity), 
while long-term \v increases.
In  contrast, the short-term \v of low mass stars is present in all stellar 
life phases while long-term variations have never been observed.

Also called the "Southern Pleiades", NGC 2516 is a rich and young open cluster 
in the constellation of Carina, that has been observed several times at
the XMM-Newton observatory, thereby allowing us the study of X-ray variability
properties on multiple time scales.
At a distance  of less than 400 pc (387 pc, \citealt{Jef97}; 346 pc, 
\citealt{Robi99}), the  cluster contains about 1300 known members spanning 
all spectral types and enabling  simultaneous study of the different processes 
driving X-ray emission from stars with different internal structures.
The metallicity of the cluster is controversial; several photometric studies 
have suggested a metal underabundance by a factor of 2 with respect to the 
Sun \citep[e.g.][]{Jef97, Jef98, Jeff01, Pin98}, while \citet{Ten02}  have 
found  solar-like metal abundance on the basis of high-resolution spectra.
Since metallicity affects the depth of the convection zone, which in turn 
should influence the dynamo's efficiency, observable differences in the
X-ray emission levels may be expected between solar and sub-solar abundances 
\citep{Mice00}.
The ROSAT  observations have shown that the G and K members in \N  were 
underactive in X-rays with respect to the Pleiades G and K stars, while 
no difference was found for M stars.  As discussed by \citet{Jef97} and
\citet{Mice00}, these results could be attributed to a low-metallicity
effect, although when taking the recent optical data into account \citep{Ten02} 
these explanations are questioned.   
Also a recent  analysis of summed XMM/EPIC data \citep{Pil05} shows that 
late-types stars in \N are significantly less luminous than those of the 
Pleiades.

\N has been observed several times with Chandra and the analysis of the 
observations was made by \citet{Har01} and \citet{Dam03}. \citet{Wo04}  
used Chandra data for a timing analysis of \N stars, finding that the 
stochastic \v  rate is similar for all sources in their sample, while the 
time scale of \v is shorter for later-type stars.

In this paper we present a study of the X-ray variability  of the NGC 2516 
stars analysing six XMM-Newton/EPIC observations spanning  $\sim$ 19 months.
We explore X-ray variability properties on short (hours), medium (months),
and long (years) time scales, we compare our data with those obtained for 
the coeval Pleiades, and we search for spectral variability.

The structure of the paper is the following:  Sect. 2 describes the  
XMM-Newton/EPIC observation set  and the data  analysis; 
Sect. 3 and 4 present the time and the spectral  analysis, respectively; 
Sect. 5 summarises the main results.

   \section{X-ray observations and data analysis}
  
\N was  observed many times during the first two years of satellite 
calibration operations.
We used only EPIC-pn data detector, since the MOS data have a lower 
statistic than pn.
The characteristic of the six EPIC/pn observations  are summarised 
in Table \ref{Tab1}.
The progressive letter in column 1 is a reference to Table \ref{t2}. 
The first two observations were centred at  7:58:20, -60:52:13 (J2000) 
and the remaining at  119.58:22, -60:45:36 (J2000). All six observations 
that we consider were performed with the thick filter.
The data span 19 months with exposure time ranging from 6 ks to $\sim$22 ks. 
All pn data were processed using the XMM-Newton Science Analysis System
(SAS) 6.0.0.
We used the {\em epchain} task  to process the EPIC/pn observation data
file, obtaining six lists with time, position, and energy of the events
recorded in the pn detector. To minimise the background due to non-X-ray 
events, we retained  only single, double, triple, and quadruple pixel events
in the 0.3-5.5 keV band.  We limited the energy band to 0.3-5.5 keV, 
since data below 0.3 keV are contaminated by low-energy electronic noise
events, while  background counts  dominate above 5.5 keV  for  coronal sources 
\citep[e.g.][]{Rea03}. 
Furthermore, we filtered the data to maximise the signal-to-noise ratio
and  to minimise the so-called proton flare phenomenon, which produces an
enhancement of noise due to protons "focused" by XMM-Newton mirrors  and
essentially indistinguishable from bona-fide X-ray events. To this end  
we applied a technique developed at  INAF - Osservatorio Astronomico of 
Palermo \citep{Scio02} that maximises the statistical significance of weak 
sources by identifying and removing fractions of the exposure time strongly
affected by  high-background episodes.

    \subsection{Source detection}

Source detection and X-ray photometry were obtained  in the 0.3 - 5.5 keV 
bandpass using the  wavelet detection code developed at the INAF- Osservatorio  
Astronomico di Palermo and based on the algorithm previously developed for the 
ROSAT/PSPC \citep{Da97,Dab97} and adapted to the XMM-Newton case. Large sets of
simulations of pure background signal were performed  to derive the appropriate 
detection threshold that limits the number of spurious detections.
We adopted a threshold that statistically retains only one spurious source per 
field. \\
An exposure map for each observation was created with the SAS task {\em eexpmap}
to perform this analysis.
For each of the six observations, we have found a number of X-ray sources  
ranging from 95 (Obs. {\em e} in Table 1) to 184 (Obs. {\em c} in Table 1). 
In order to find the X-ray  multiply observed sources, we cross-matched the 
detections of each observation with all the others  adopting a threshold of 
20${\arcsec}$, derived by taking the formal error of the detected source 
positions into account. However we found that more than 90\% of the 
cross-matchs are within 8${\arcsec}$.
We also found that 42 sources were detected six times, 23 five times, 
39 four time, 40 three times, 60 two times, and 99 just one.

\begin{table*}
\centering
\caption{XMM-Newton/EPIC/pn  observations of NGC 2516}. 
\bigskip
\begin{tabular}{clcclcccc}
\hline
\hline
\multicolumn{1}{c}{Obs.} &\multicolumn{1}{c}{Obs. Id} &\multicolumn{1}{c}{Orbit}&
\multicolumn{1}{c}{Filtered}&\multicolumn{1}{c}{Start}& \multicolumn{1}{c}{Nr. of X-ray} &
\multicolumn{1}{c}{Nr. of detected}& \multicolumn{1}{c}{Nr. of optical}  \\
\multicolumn{1}{c}{} & & \multicolumn{1}{c}{Nr.}
&\multicolumn{1}{c}{Time [ks]} &\multicolumn{1}{c}{Date} &
\multicolumn{1}{c}{sources}&\multicolumn{1}{c}{members} &\multicolumn{1}{c}{members}  \\
\hline
{\em a} & 0113891001 & 060 &19.8 &  2000 April 06 &128 & 87 &409\\
{\em b} & 0113891101 & 060 &14.3&   2000 April 07 & 118& 73  &408\\
{\em c} & 0126511201 & 092 & 22.3  & 2000 June 10 & 184& 128 &419 \\
{\em d} & 0134531201 & 209 &18.8&  2001 January 29 & 174& 121 &408\\
{\em e} & 0134531301 & 209 &~6.0&  2001 January 29 & 95& 68 &402\\
{\em f} & 0134531501 & 346 &18.5  & 2001 October 29 & 168& 112 &423 \\
\hline
\end{tabular}
\label{Tab1}
\end{table*}

\begin{table*}
\caption{X-ray and optical properties of NGC 2516 members in the XMM-Newton/EPIC/pn
observations having more than 25 counts.} 
\bigskip
\scriptsize
\begin{tabular}{clcclccccccc}
\hline
\hline
\multicolumn{1}{c}{JTH} &\multicolumn{1}{c}{Obs.} &\multicolumn{1}{c}{Ra}
&\multicolumn{1}{c}{Dec}& \multicolumn{1}{c}{V}&\multicolumn{1}{c}{B-V}&
\multicolumn{1}{c}{V-I} & \multicolumn{1}{c}{Exposure} &\multicolumn{1}{c}{Rate $\pm$ err}
&\multicolumn{1}{c}{Results of}  \\
&  &\multicolumn{1}{c}{J2000} &\multicolumn{1}{c}{J2000}  &  &  & &\multicolumn{1}{c}{[sec]}
&\multicolumn{1}{c}{cnt/ks} & \multicolumn{1}{c}{the K-S test}  \\
\hline
  323 &   {\em f} & 07:56:22.80  & -60:51:42.7 & 11.556 &  0.513 &   0.659 &  4894.6 & 16.750 $\pm$ 1.850 & -\\
15496 &   {\em a} & 07:56:46.30  & -60:48:59.5 &  8.360 &  0.020 &    &  8094.8 &  4.200 $\pm$  0.720 & -\\
15496 &   {\em b} & 07:56:46.30  & -60:48:59.5 &  8.360 &  0.020 &    &  5880.3 &  5.100 $\pm$  0.931 & -\\
15496 &   {\em c} & 07:56:46.30  & -60:48:59.5 &  8.360 &  0.020 &    &  8806.8 &  4.825 $\pm$  0.740 & -\\
15496 &   {\em d} & 07:56:46.30  & -60:48:59.5 &  8.360 &  0.020 &    &  6764.2 &  5.915 $\pm$  0.935 &90\%- 99\%\\
15496 &   {\em f} & 07:56:46.30  & -60:48:59.5 &  8.360 &  0.020 &    &  6030.5 &  5.305 $\pm$  0.938 & -\\
\end{tabular}
  \end{table*}
  \begin{table*}
  \addtocounter{table}{-1}
  \caption{Continued}
  \bigskip
  \scriptsize
  \label{t2}
  \begin{tabular}{clcclccccccc}
  \hline\hline
  \multicolumn{1}{c}{JTH} &\multicolumn{1}{c}{Obs.} &\multicolumn{1}{c}{Ra$_{ott}$}
&\multicolumn{1}{c}{Dec$_{ott}$
}& \multicolumn{1}{c}{V}&\multicolumn{1}{c}{B-V}& \multicolumn{1}{c}{V-I} & \multicolumn{1}{c}{Exposure}
&\multicolumn{1}{c}{Rate $\pm$ err} &\multicolumn{1}{c}{K-S}  \\
  &  &\multicolumn{1}{c}{J2000} &\multicolumn{1}{c}{J2000}  &  &  & &\multicolumn{1}{c}{[sec]}
&\multicolumn{1}{c}{
cnt/ks} &\multicolumn{1}{c}{} & \multicolumn{1}{c}{}&\multicolumn{1}{c}{}  \\
  \hline
    522 & {\em e} & 07:57:20.62   & -60:55:46.4   & 17.301 &  1.476 &	2.020 &  2240.6 & 11.160 $\pm$  2.230 & -\\
    523 & {\em c} & 07:57:20.86   & -60:44: 0.8   & 14.945 &  1.065 &	1.204 & 13991.0 &  3.790 $\pm$  0.520 & -\\
    523 & {\em d} & 07:57:20.86   & -60:44: 0.8   & 14.945 &  1.065 &	1.204 & 10000.0 &  7.200 $\pm$  0.850 & -\\
    523 & {\em f} & 07:57:20.86   & -60:44: 0.8   & 14.945 &  1.065 &	1.204 &  9982.4 &  6.410 $\pm$  0.800 & -\\
    523 & {\em a} & 07:57:20.86   & -60:44: 0.8   & 14.945 &  1.065 &	1.204 &  8317.3 &  7.570 $\pm$  0.950 & $\geq$99\%\\
    523 & {\em b} & 07:57:20.86   & -60:44: 0.8   & 14.945 &  1.065 &	1.204 &  6039.3 &  6.460 $\pm$  1.030 & -\\
\end{tabular}
  \end{table*}
  \begin{table*}
  \addtocounter{table}{-1}
  \caption{Continued}
  \bigskip
  \scriptsize
  \label{t2}
  \begin{tabular}{clcclccccccc}
  \hline\hline
  \multicolumn{1}{c}{JTH} &\multicolumn{1}{c}{Obs.} &\multicolumn{1}{c}{Ra$_{ott}$}
&\multicolumn{1}{c}{Dec$_{ott}$ }& \multicolumn{1}{c}{V}&\multicolumn{1}{c}{B-V}& \multicolumn{1}{c}{V-I} & \multicolumn{1}{c}{Exposure}
&\multicolumn{1}{c}{Rate $\pm$ err} &\multicolumn{1}{c}{K-S}  \\
  &  &\multicolumn{1}{c}{J2000} &\multicolumn{1}{c}{J2000}  &  &  & &\multicolumn{1}{c}{[sec]}
&\multicolumn{1}{c}{
cnt/ks} &\multicolumn{1}{c}{} & \multicolumn{1}{c}{}&\multicolumn{1}{c}{}  \\
  \hline
  15501 & {\em c} & 07:57:47.69  & -60:36:35.6   &  8.390 & -0.010 &	& 11316.0 &  6.800 $\pm$  0.780 & -\\
  15501 & {\em f} & 07:57:47.69  & -60:36:35.6   &  8.390 & -0.010 &	&  8449.1 &  6.860 $\pm$  0.900 & -\\
    625 & {\em c} & 07:57:51.22  & -60:42:11.9   & 17.771 &  1.589 &   2.120 & 17025.0 &  2.640 $\pm$  0.390 & -\\
    624 & {\em c} & 07:57:51.50  & -60:43:17.0   & 18.702 &   &   2.568 & 17728.0 &  6.710 $\pm$  0.620 & 90\%-99\%\\
    624 & {\em f} & 07:57:51.50  & -60:43:17.0   & 18.702 &   &   2.568 & 13414.0 &  5.220 $\pm$  0.620 & -\\
    635 & {\em f} & 07:57:53.38  & -60:46:13.9   & 19.713 &   &   2.866 & 13631.0 &  6.900 $\pm$  0.710 & $\geq$99\%\\
\end{tabular}
  \end{table*}
  \begin{table*}
  \addtocounter{table}{-1}
  \caption{Continued}
  \bigskip
  \scriptsize
  \label{t2}
  \begin{tabular}{clcclccccccc}
  \hline\hline
  \multicolumn{1}{c}{JTH} &\multicolumn{1}{c}{Obs.} &\multicolumn{1}{c}{Ra$_{ott}$}
&\multicolumn{1}{c}{Dec$_{ott}$
}& \multicolumn{1}{c}{V}&\multicolumn{1}{c}{B-V}& \multicolumn{1}{c}{V-I} & \multicolumn{1}{c}{Exposure}
&\multicolumn{1}{c}{Rate $\pm$ err} &\multicolumn{1}{c}{K-S}  \\
  &  &\multicolumn{1}{c}{J2000} &\multicolumn{1}{c}{J2000}  &  &  & &\multicolumn{1}{c}{[sec]}
&\multicolumn{1}{c}{
cnt/ks} &\multicolumn{1}{c}{} & \multicolumn{1}{c}{}&\multicolumn{1}{c}{}  \\
  \hline
    676 & {\em d} & 07:58:  4.68   & -60:52:42.7  & 18.790 &  1.602 &	2.442 & 11801.0 &  5.850 $\pm$  0.700 & $\geq$99\%\\
    676 & {\em f} & 07:58:  4.68   & -60:52:42.7  & 18.790 &  1.602 &	2.442 & 10287.0 &  3.890 $\pm$  0.610 & -\\
    679 & {\em c} & 07:58:  5.40   & -60:46:14.2  & 14.035 &  1.006 &	1.124 & 19333.0 &  5.070 $\pm$  0.510 & 90\%-99\%\\
    679 & {\em d} & 07:58:  5.40   & -60:46:14.2  & 14.035 &  1.006 &	1.124 & 15578.0 &  3.020 $\pm$  0.440 & -\\
    679 & {\em a} & 07:58:  5.40   & -60:46:14.2  & 14.035 &  1.006 &	1.124 & 12181.0 & 15.020 $\pm$  1.110 & $\geq$99\%\\
    679 & {\em b} & 07:58:  5.40   & -60:46:14.2  & 14.035 &  1.006 &	1.124 &  8864.7 &  6.660 $\pm$  0.870 & -\\
\end{tabular}
  \end{table*}
  \begin{table*}
  \addtocounter{table}{-1}
  \caption{Continued}
  \bigskip
  \scriptsize
  \label{t2}
  \begin{tabular}{clcclccccccc}
  \hline\hline
  \multicolumn{1}{c}{JTH} &\multicolumn{1}{c}{Obs.} &\multicolumn{1}{c}{Ra$_{ott}$}
&\multicolumn{1}{c}{Dec$_{ott}$
}& \multicolumn{1}{c}{V}&\multicolumn{1}{c}{B-V}& \multicolumn{1}{c}{V-I} & \multicolumn{1}{c}{Exposure}
&\multicolumn{1}{c}{Rate $\pm$ err} &\multicolumn{1}{c}{K-S}  \\
  &  &\multicolumn{1}{c}{J2000} &\multicolumn{1}{c}{J2000}  &  &  & &\multicolumn{1}{c}{[sec]}
&\multicolumn{1}{c}{
cnt/ks} &\multicolumn{1}{c}{} & \multicolumn{1}{c}{}&\multicolumn{1}{c}{}  \\
  \hline
    735 & {\em b} & 07:58:19.90  & -60:42: 0.2   & 12.336 &  0.583 &   0.709 &  5925.9 & 10.630 $\pm$  1.340 & -\\
    735 & {\em e} & 07:58:19.90  & -60:42: 0.2   & 12.336 &  0.583 &   0.709 &  4277.9 &  7.250 $\pm$  1.300 & -\\
    737 & {\em c} & 07:58:20.16  & -60:41:38.7   & 12.871 &  0.660 &   0.760 & 17697.0 &  2.030 $\pm$  0.340 & -\\
    737 & {\em d} & 07:58:20.16  & -60:41:38.7   & 12.871 &  0.660 &   0.760 & 13716.0 &  8.310 $\pm$  0.780 & -\\
    737 & {\em f} & 07:58:20.16  & -60:41:38.7   & 12.871 &  0.660 &   0.760 & 14664.0 &  9.750 $\pm$  0.820 & -\\
    737 & {\em e} & 07:58:20.16  & -60:41:38.7   & 12.871 &  0.660 &   0.760 &  4277.9 &  7.250 $\pm$  1.300 & -\\
\end{tabular}
  \end{table*}
  \begin{table*}
  \addtocounter{table}{-1}
  \caption{Continued}
  \bigskip
  \scriptsize
  \label{t2}
  \begin{tabular}{clcclccccccc}
  \hline\hline
  \multicolumn{1}{c}{JTH} &\multicolumn{1}{c}{Obs.} &\multicolumn{1}{c}{Ra$_{ott}$}
&\multicolumn{1}{c}{Dec$_{ott}$
}& \multicolumn{1}{c}{V}&\multicolumn{1}{c}{B-V}& \multicolumn{1}{c}{V-I} & \multicolumn{1}{c}{Exposure}
&\multicolumn{1}{c}{Rate $\pm$ err} &\multicolumn{1}{c}{K-S}  \\
  &  &\multicolumn{1}{c}{J2000} &\multicolumn{1}{c}{J2000}  &  &  & &\multicolumn{1}{c}{[sec]}
&\multicolumn{1}{c}{
cnt/ks} &\multicolumn{1}{c}{} & \multicolumn{1}{c}{}&\multicolumn{1}{c}{}  \\
  \hline
    776 & {\em d} & 07:58:31.66 & -60:53:12.0  & 17.870 &  1.521 &   2.306 & 11911.0 &  3.860 $\pm$  0.569 & -\\
    776 & {\em f} & 07:58:31.66 & -60:53:12.0  & 17.870 &  1.521 &   2.306 & 10101.0 &  3.020 $\pm$  0.547 & -\\
    776 & {\em a} & 07:58:31.66 & -60:53:12.0  & 17.870 &  1.521 &   2.306 & 16396.0 &  4.330 $\pm$  0.514 & -\\
    776 & {\em e} & 07:58:31.66 & -60:53:12.0  & 17.870 &  1.521 &   2.306 &  3802.3 &  4.075 $\pm$  1.035 & -\\
    780 & {\em c} & 07:58:31.66 & -60:53:12.0  & 14.762 &  0.996 &   1.139 & 12055.0 &  2.985 $\pm$  0.498 & -\\
    780 & {\em d} & 07:58:31.66 & -60:53:12.0  & 14.762 &  0.996 &   1.139 & 11911.0 &  3.860 $\pm$  0.569 & -\\
\end{tabular}
  \end{table*}
  \begin{table*}
  \addtocounter{table}{-1}
  \caption{Continued}
  \bigskip
  \scriptsize
  \label{t2}
  \begin{tabular}{clcclccccccc}
  \hline\hline
  \multicolumn{1}{c}{JTH} &\multicolumn{1}{c}{Obs.} &\multicolumn{1}{c}{Ra$_{ott}$}
&\multicolumn{1}{c}{Dec$_{ott}$
}& \multicolumn{1}{c}{V}&\multicolumn{1}{c}{B-V}& \multicolumn{1}{c}{V-I} & \multicolumn{1}{c}{Exposure}
&\multicolumn{1}{c}{Rate $\pm$ err} &\multicolumn{1}{c}{K-S}  \\
  &  &\multicolumn{1}{c}{J2000} &\multicolumn{1}{c}{J2000}  &  &  & &\multicolumn{1}{c}{[sec]}
&\multicolumn{1}{c}{
cnt/ks} &\multicolumn{1}{c}{} & \multicolumn{1}{c}{}&\multicolumn{1}{c}{}  \\
  \hline
    820 & {\em f} & 07:58:41.78 & -60:50:52.0   & 17.733 &  1.564 &   2.123 & 12899.0 &  4.650 $\pm$  0.600 & -\\
    820 & {\em b} & 07:58:41.78 & -60:50:52.0   & 17.733 &  1.564 &   2.123 & 10667.0 &  3.660 $\pm$  0.590 & -\\
    823 & {\em f} & 07:58:42.84 & -60:40:20.4   & 13.641 &  0.781 &   0.851 & 13214.0 &  1.970 $\pm$  0.390 & -\\
    826 & {\em c} & 07:58:43.39 & -60:55:26.8   & 13.840 &  0.818 &   0.963 &  9528.0 & 16.900 $\pm$  1.330 & -\\
    826 & {\em d} & 07:58:43.39 & -60:55:26.8   & 13.840 &  0.818 &   0.963 &  9315.6 & 28.120 $\pm$  1.740 & 90\%-99\%\\
    826 & {\em f} & 07:58:43.39 & -60:55:26.8   & 13.840 &  0.818 &   0.963 &  8385.0 & 66.070 $\pm$  2.810 & $\geq$99\%\\
\end{tabular}
  \end{table*}
  \begin{table*}
  \addtocounter{table}{-1}
  \caption{Continued}
  \bigskip
  \scriptsize
  \label{t2}
  \begin{tabular}{clcclccccccc}
  \hline\hline
  \multicolumn{1}{c}{JTH} &\multicolumn{1}{c}{Obs.} &\multicolumn{1}{c}{Ra$_{ott}$}
&\multicolumn{1}{c}{Dec$_{ott}$
}& \multicolumn{1}{c}{V}&\multicolumn{1}{c}{B-V}& \multicolumn{1}{c}{V-I} & \multicolumn{1}{c}{Exposure}
&\multicolumn{1}{c}{Rate $\pm$ err} &\multicolumn{1}{c}{K-S}  \\
  &  &\multicolumn{1}{c}{J2000} &\multicolumn{1}{c}{J2000}  &  &  & &\multicolumn{1}{c}{[sec]}
&\multicolumn{1}{c}{
cnt/ks} &\multicolumn{1}{c}{} & \multicolumn{1}{c}{}&\multicolumn{1}{c}{}  \\
  \hline
    868 & {\em b} & 07:58:56.50  & -60:47:22.3  & 17.183 &  1.492 &   1.930 &  8064.1 &  4.840 $\pm$  0.770 & -\\
    870 & {\em c} & 07:58:57.17  & -60:36:13.6  & 14.867 &  1.034 &   1.171 &  9930.8 & 10.880 $\pm$  1.050 & -\\
    870 & {\em d} & 07:58:57.17  & -60:36:13.6  & 14.867 &  1.034 &   1.171 &  7865.9 &  6.610 $\pm$  0.920 & -\\
    870 & {\em f} & 07:58:57.17  & -60:36:13.6  & 14.867 &  1.034 &   1.171 &  8677.2 & 10.030 $\pm$  1.070 & -\\
    871 & {\em a} & 07:58:57.74  & -60:43: 3.6  & 14.760 &  0.990 &   1.140 &  7847.0 & 26.700 $\pm$ 1.845 & $\geq$99\%\\
    871 & {\em c} & 07:58:57.74  & -60:43: 3.6  & 14.760 &  0.990 &   1.140 & 15135.0 &  9.975 $\pm$  0.812 & 90\%-99\%\\
\end{tabular}
  \end{table*}
  \begin{table*}
  \addtocounter{table}{-1}
  \caption{Continued}
  \bigskip
  \scriptsize
  \label{t2}
  \begin{tabular}{clcclccccccc}
  \hline\hline
  \multicolumn{1}{c}{JTH} &\multicolumn{1}{c}{Obs.} &\multicolumn{1}{c}{Ra$_{ott}$}
&\multicolumn{1}{c}{Dec$_{ott}$
}& \multicolumn{1}{c}{V}&\multicolumn{1}{c}{B-V}& \multicolumn{1}{c}{V-I} & \multicolumn{1}{c}{Exposure}
&\multicolumn{1}{c}{Rate $\pm$ err} &\multicolumn{1}{c}{K-S}  \\
  &  &\multicolumn{1}{c}{J2000} &\multicolumn{1}{c}{J2000}  &  &  & &\multicolumn{1}{c}{[sec]}
&\multicolumn{1}{c}{
cnt/ks} &\multicolumn{1}{c}{} & \multicolumn{1}{c}{}&\multicolumn{1}{c}{}  \\
  \hline
    967 & {\em c} & 07:59:28.97  & -60:39:36.8  & 17.410 &  1.520 &   2.006 &  9534.9 &  2.940 $\pm$  0.550 & $\geq$99\%\\
    977 & {\em c} & 07:59:32.04  & -60:48:47.2  & 13.879 &  0.834 &   1.019 &  9887.9 & 27.510 $\pm$  1.670 & 90\%-99\%\\
    977 & {\em d} & 07:59:32.04  & -60:48:47.2  & 13.879 &  0.834 &   1.019 &  9696.1 & 22.380 $\pm$  1.520 & 90\%-99\%\\
    977 & {\em f} & 07:59:32.04  & -60:48:47.2  & 13.879 &  0.834 &   1.019 & 10348.0 & 20.680 $\pm$  1.410 & -\\
    977 & {\em a} & 07:59:32.04  & -60:48:47.2  & 13.879 &  0.834 &   1.019 &  8466.0 & 28.470 $\pm$  1.830 & -\\
    977 & {\em b} & 07:59:32.04  & -60:48:47.2  & 13.879 &  0.834 &   1.019 &  6154.6 & 27.620 $\pm$  2.120 & -\\
\hline
  \end{tabular}
  \label{t2}
  \end{table*}

\subsection{Cluster members and identifications}
We  compiled an optical catalog of cluster members based on the list of 
1254 stars compiled by \citet{Jeff01} and based on B, V, and I photometry.
Furthermore we appended 43 cluster stars brighter than V$=$ 9.8 that were 
not present in the former sample.
The number of X-ray sources identified as optical members are reported in 
column 7 of Table 1 for each of the six observations; the number of the 
optical members in each observation is reported in column 8.
The cross-identification between X-ray and optical sources for each of the 
six observations was made  in two steps: in the first  we  searched for  a 
systematic offset between X-ray source positions and the optical position 
of the members.  In the second, we  corrected the X-ray positions for this 
systematic offset, before matching the X-ray and optical member positions, 
and then retained an identification if the offset between X-ray and optical 
positions was less than 8${\arcsec}$.
The choice of such  a limiting distance is a good compromise between the attempt
to minimise the number of spurious identifications 
for an offset 
bona-fide optical  counterparts.
For 8 cases it was impossible to resolve very close stars, owing to the limited 
spatial resolution  of the X-ray telescope.
For these unresolved sources, the X-ray flux was  divided evenly between 
the optical candidates in the absence of more information.

Following \citet{Dam03}, we attribute spectral types that use B-V and V-I 
optical colours corrected for the average cluster reddening E(B-V) = 0.12. 
Detected members cover the whole mass range present in the optical catalog. 
Among these stars we considered in the following \v studies all those sources
with more than 25 total counts in a single pn observation, for a total of 
474 detections.  Table 2 summarise their optical and X-ray characteristics.

       \section{Time variability}
The analysis of all observations of NGC 2516 allow us to explore the X-ray
\v on time scales that range from hours to 19 months. We obtained light curves
in the 0.3-5.5 keV band for all detected X-ray sources by extracting the 
photon arrival time within circular regions selected interactively using
Astronomical Data Visualization DS9 display software and subtracted from an
area-scaled background. We adopted a radius of 3.5 times the radius determined 
by the PWDetect algorithm to select sources and background regions. 
In some cases we  used a smaller radius to exclude contributions from nearby
stars.
In general, detections have relatively low statistics with 90\% of the 
detections having less than 190 counts and only 23 detections more than 
300 counts.

In order to have a statistical evaluation of the X-ray variability, we 
applied the unbinned Kolmogorov-Smirnov (K-S) test to the X-ray photon time
series of our detected sources for each observation.
Column 10 of  Table 2 reports the results in terms of the confidence level 
at which we can reject the hypothesis that the source in the given observation
is constant.
Since the exposure times of the observations are in the 6 - 22.3 ksec range, 
we will refer to this analysis as short-time- scale variability.
For each observation, we also ran the K-S test on the counts detected in 
the background regions to monitor possible background variability. 
In the few cases in which the background counts are variable with a confidence
level (CL) $>$ 99$\%$, the background counts  are much lower than those
of  the source, making us confident of the results of the test variability.

\begin{table*}
\centering
\caption{Results of the K-S test for each X-ray detected source.}
\bigskip
\begin{tabular}{cccccc}
\hline
\hline
\multicolumn{1}{c}{Confidence$^1$} &\multicolumn{1}{c}{Number of total} &
\multicolumn{1}{c}{Number of cluster}&\multicolumn{1}{c}{Number of non members} \\
\multicolumn{1}{c}{level} & \multicolumn{1}{c}{sources} &\multicolumn{1}{c}{members} & & &   \\
\hline
$<$90\% & 490 (74\%) &345 (73\%) &145 (79\%)  &  \\
90\%-99\% &96 (15\%)  &72 (15\%)& 24 (13\%) &  \\
$\geq$ 99\% & 71 (11\%) & 57 (12\%) & 14 (8\%) & \\
\hline
\end{tabular}
\label{ksT}

$^1$Confidence level for the rejection of the constant source hypothesis.
\end{table*}

\begin{table*}
\centering
\caption{Results of the K-S test for the stars of  NGC 2516 grouped by spectral type.}
\bigskip
\begin{tabular}{clccrccc}
\hline
\hline
\multicolumn{1}{c}{Confidence$^1$} && &\multicolumn{1}{c}{Number of cluster members } &&&\\
\multicolumn{1}{c}{level} &\multicolumn{1}{c}{B} & \multicolumn{1}{c}{dA}&\multicolumn{1}{c}{dF} &\multicolumn{1}{c}{dG} & \multicolumn{1}{c}{dK} & \multicolumn{1}{c}{dM}  \\
\hline
$<$90\% & 9 (64\%) &62 (74\%) &9 (69\%)  & 109 (83\%) & 119 (70\%)  &37 (59\%) & \\
90\%-99\% &2 (14\%)  &13 (15\%)& 0  & 14 (11\%) & 32 (19\%) & 11 (17\%) \\
$\geq$ 99\% & 3 (22\%) & 9 (11\%) & 4(31\%) &8 (6\%)  &18 (11\%) &15 (24\%) \\
\hline
\end{tabular}
\label{Tab3}

$^1$Confidence level for the rejection of the constant source hypothesis.
\end{table*}

\begin{figure}
\centerline{\psfig{figure=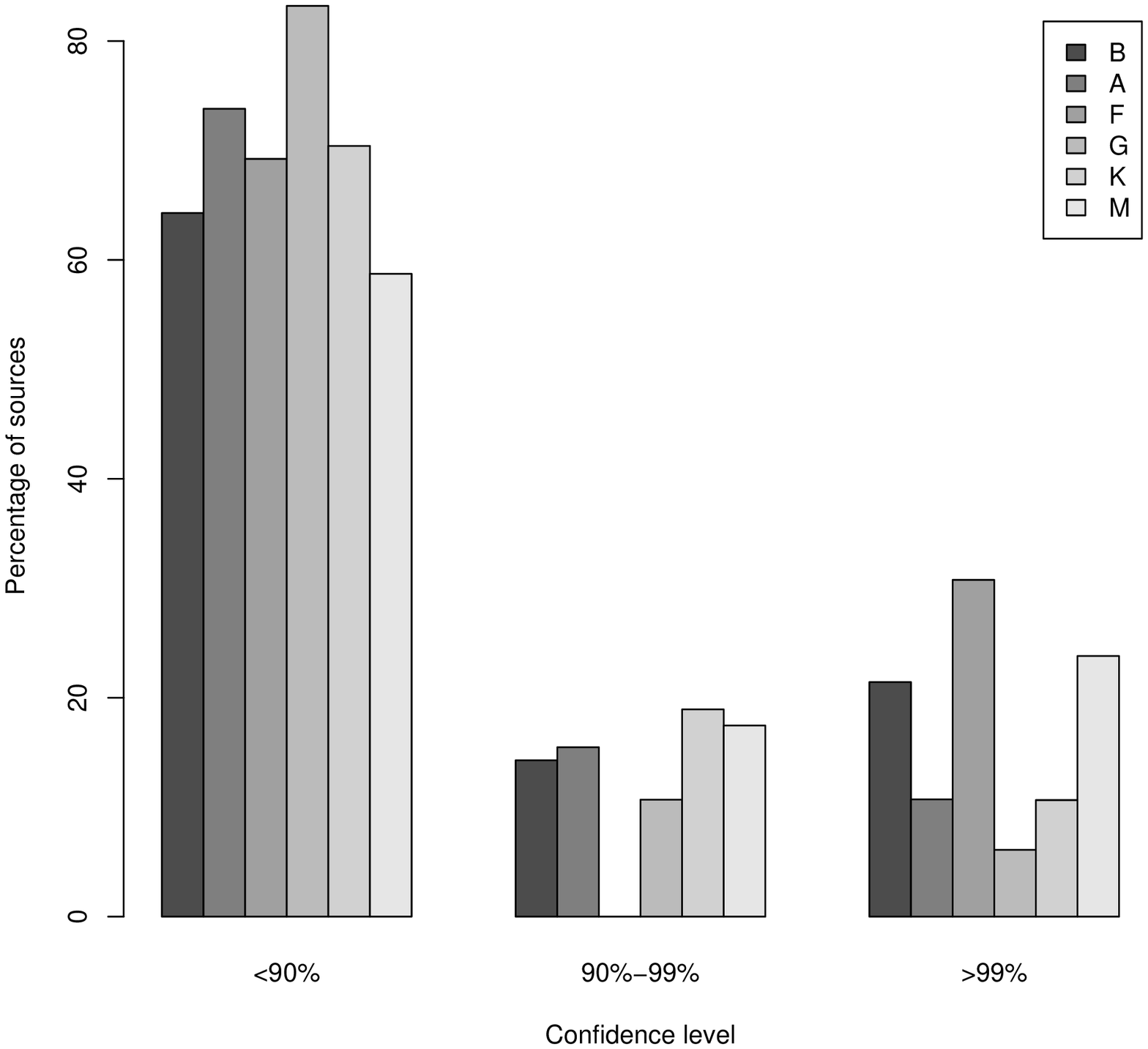,width=8cm,angle=-90}}
\caption{Percentage of detected stars vs. \v  confidence level and spectral type.}
\label{ks}
\end{figure}

Table 3 summarises the  K-S test results for the total  sample of X-ray sources 
(column 2), for the members (column 3), and for  non-members (column 4).  
Among cluster  members, only a small fraction (12\%) of stars are variable with 
a CL $\geq$99\%, while 73\% are not variable (CL $<$90\%); variability on a 
short-time-scale is not very common in NGC 2516, at least  given the statistics 
of our observations. 
Table 4 and Figure \ref{ks} summarise the K-S test results for member stars of
different spectral types; dF stars are the sources with the highest rate of 
variability (although the sample is very small), dG are those that are less 
variable. In contrast, \citet{Wo04} find that the dF and B in their sample have 
the lowest  rate of variability. 
Frequency of the \v on a short-time-scale is approximately  equal among late- and
early-types stars.   
The fraction of sources with  significant \v is very  small for each spectral type
suggesting that, on short time scales, the properties of \v do not depend strongly 
on the mass.
However, the sensitivity to \v depends on count statistics, and the absence of 
\v in faint stars  may be not intrinsic but instead due to low counting statistics.
Also the comparisons of our results with those obtained for other sample stars 
are limited by different statistics.

    \subsection{Time X-ray distribution functions}

\begin{figure}
\centerline{\psfig{figure=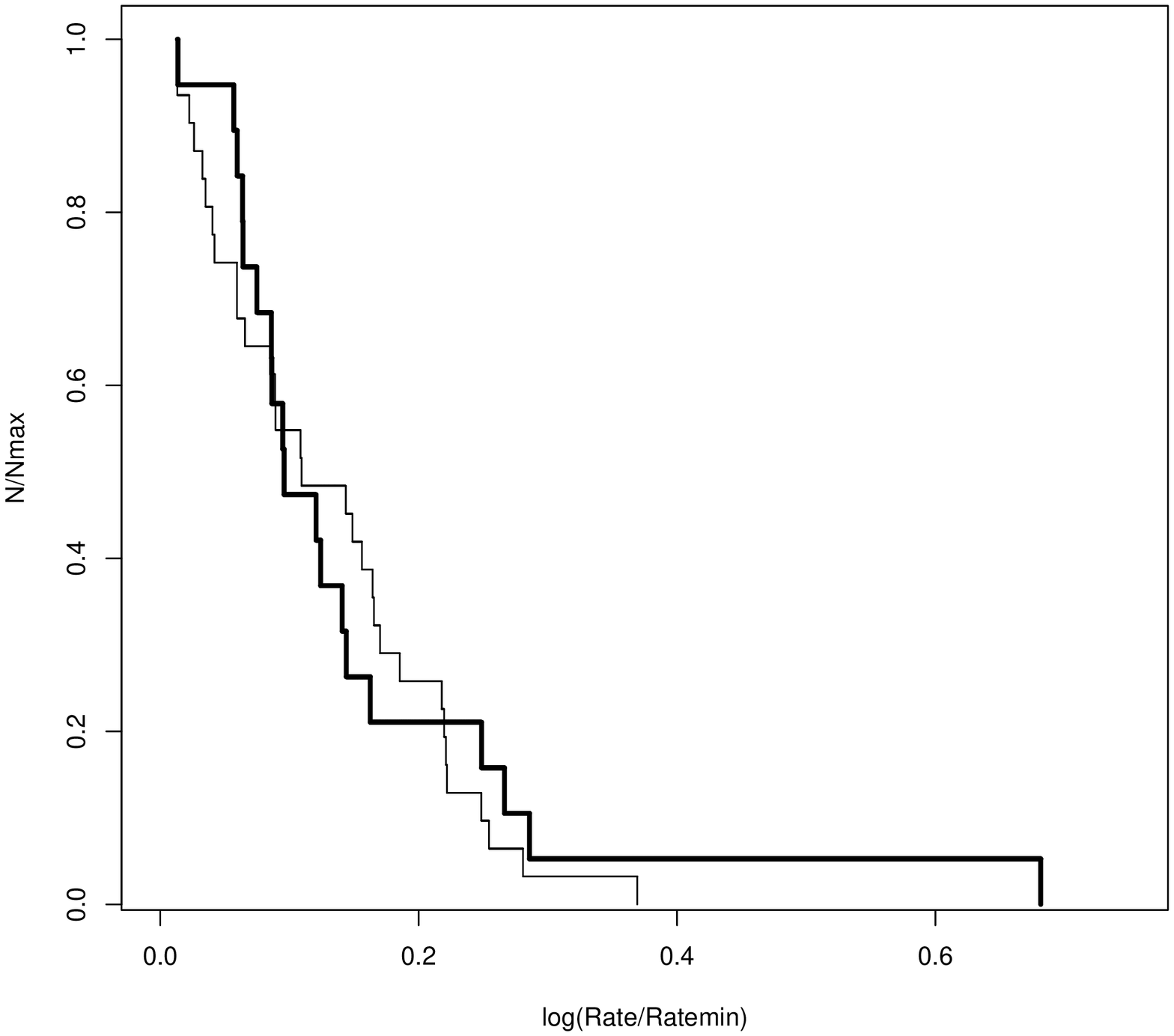,angle=-90,width=9cm}}
\caption{The Time XAD for dF7-dK2 stars on short ($\leq $ 1 day, Obs. 
{\em a-b, d-e}, thin line) and long ($\sim$ 17 months, Obs.  {\em c-f}, 
thick line) time scales.}
\label{GT}
\centerline{\psfig{figure=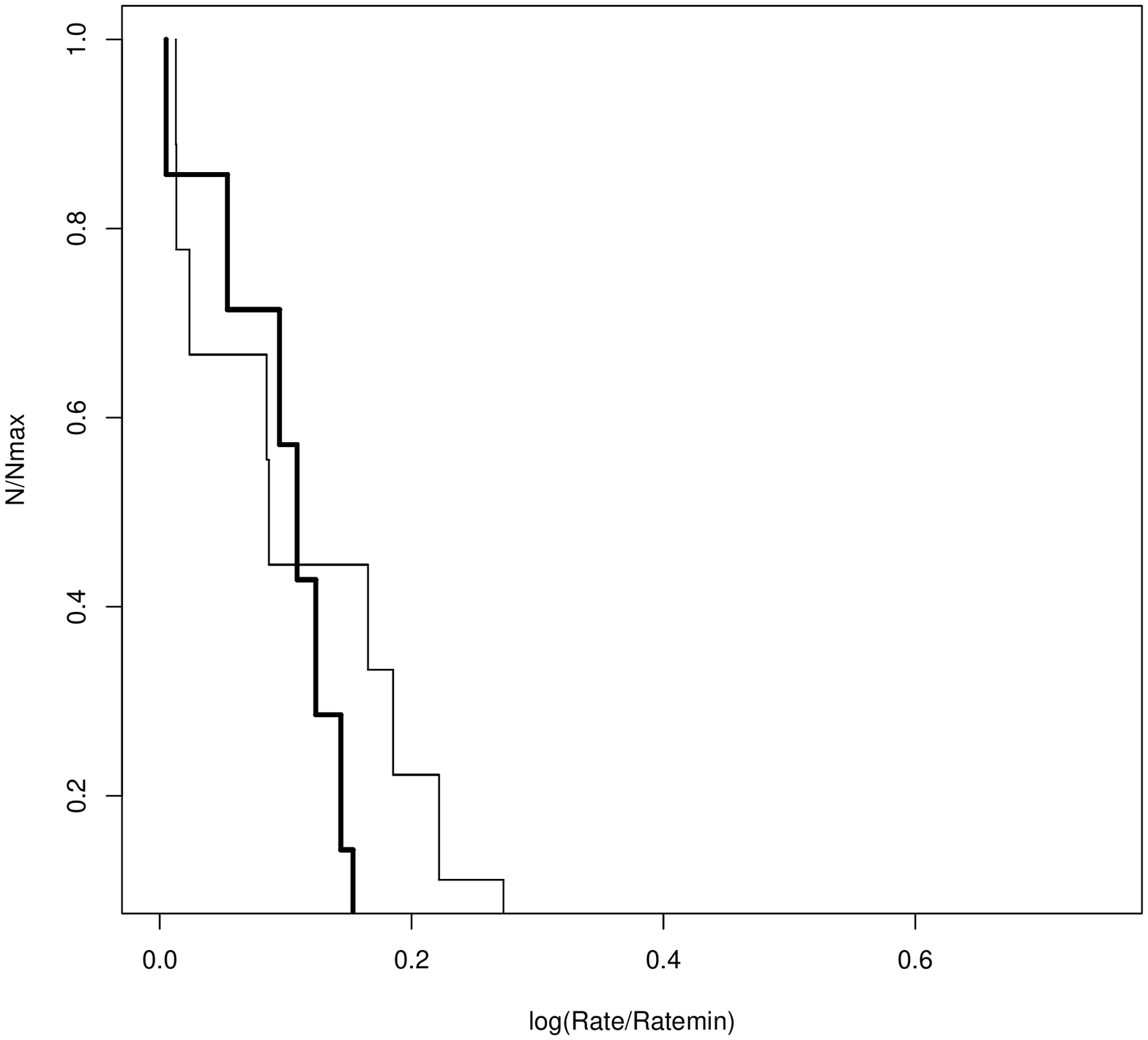,angle=-90,width=9cm}}
\caption{As in Figure \ref{GT} for  dK3-dM on short (thin line) and long (thick line) time scales.}
\label{MT}
\end{figure}

We derived the time amplitude X-ray luminosity distribution function (Time XAD)
for the dF7-dK2 (0.5 $\leq$B-V$\leq$0.99) and dK3-dM  (B-V $>$0.9) stars as in 
\citet{Mar03a}. 
Time XAD yields the fraction of time that a star spends with a count rate higher, 
by a  given factor, than its minimum value. This distribution is constructed by 
considering for each star the ratio between the count rate observed during a 
given  observation and the minimum rate for that source.
We  computed the Time XAD on a given time scale considering only observations 
obtained at a time separation of the order of the time scale we want to study.
For example, if we want to study a 1-day time scale, we can consider Obs. 
{\em a, b, d}, and {\em e} of Table 1.  In Figure \ref{GT} the Time XAD on 
short ($<$ 1 day) time scale 
with that on a medium (17 months) time scale. The same plot for dK3-dM stars is 
shown in Figure \ref{MT}.  
To explore the presence of variations on medium time scale amplitudes, we excluded
the stars  variable (CL $\geq$ 99\%) on short time scales that could produce 
spurious variability. 
Both the dF7-dK2 and dK3-dM  distributions on short and medium time scales   
appear very alike, indicating that medium-term variations, if they exist at all, 
must have a much larger  amplitude than those on short time scales. The null 
hypothesis that the XADs on short and medium time scales of dF7-dK2 stars are 
drawn from the same parent distribution cannot be rejected. 

    \subsection{Long-term \v}

The Sun's 11-year activity cycle, as observed in optical, radio, ultraviolet, 
and X-ray bands, is a well-established phenomenon.
In the soft X-ray bandpass, the flux varies by about one order of magnitude 
during the solar cycle \citep{Kre77, Pe00}.
The amplitude variations of the Sun are strongly dependent on time scale, 
with the highest probability of observing high-amplitude variations, between 
4 and 7.5 years \citep{Mice03}. On these time scales, typically the observed 
variations are on average of the order of 1 dex, but the spread is very wide 
with an 80\% probability of detecting variations in the 0.1 $-$ 1.8 dex range.
Solar-like cycles have been observed in late-type stars 
from chromospheric flux variations \citep{Wi78, Ba95}. Only in  more recent years, 
X-ray long-term \v possibly due to cyclical variations has been observed in stars 
\citep{He03,Fa04}.
By combining X-ray data obtained with ROSAT and, at present, XMM-Newton, we 
explored the long-term \v of the late-type stars of NGC 2516.

\begin{figure}
\centerline{\psfig{figure=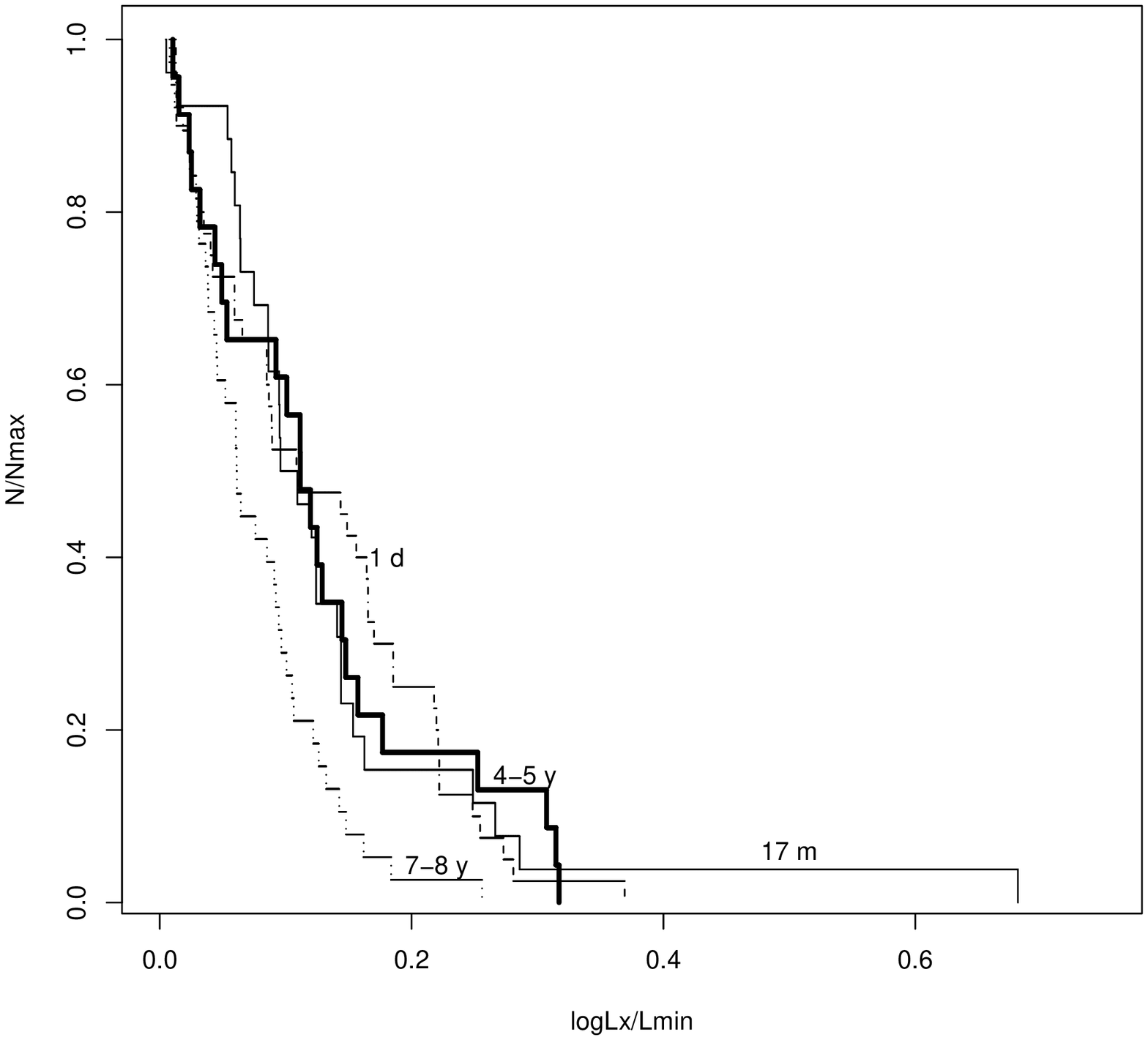,angle=-90,width=9cm}}
\caption{The Time XADs for dF7-dM on 7-8 years (dotted line), 
on 4-5 years (thick line), on $\sim$ 17 months (thin solid line) 
and 1 day (dot-dashed line) time scales.}
\label{long}
\end{figure}

Thirty-eight  late-type stars (dF7-dM) of the cluster observed with XMM-Newton 
were observed with ROSAT/PSPC in 1993 \citep{Jef97} and twenty-three  with 
ROSAT/HRI in 1997 \citep{Mice00}, allowing us to explore \v on 7-8  and 4-5 year
time scales, respectively.
In order to compare our results with ROSAT observations, we derived EPIC/pn 
X-ray luminosities in the(0.15-2.0) keV bandpass, assuming a constant 
conversion factor to convert count rates to flux, as more extensively 
described in  Sect. 3.3. 
For each star, several points were available from EPIC/pn and only one from 
PSPC (L$_{PSPC}$) and HRI (L$_{HRI}$).
We reduced the influence of short and medium \v in the EPIC/pn data, 
averaging the latter to a single mean value (L$_{pn}$).
The amplitude of \v is defined as log(L$_x$/L$_{xmin}$), where L$_x$ is the 
higher and L$_{xmin}$, the lower value between L$_{pn}$ and L$_{PSPC}$ and 
between L$_{pn}$ and L$_{HRI}$, respectively.
Figure \ref{long} shows the XADs of the stars observed both with PSPC and 
EPIC/pn observations, and with HRI and EPIC/pn. 
For comparison in the same figure, we also show the XADs  limited to 
EPIC/pn dF7-dM stars on time scales of 1 day and on 17 months. 
The XADs on time scales of 4-5 years and on 7-8 years are  marginally 
(CL $\geq$ 96\%) different, suggesting that long-term variations could 
be present.  However, long-term variations, if  they exist, must have
a smaller  amplitude than the short and medium term variations, or 
comparable to them.  
Our finding  that there is no evidence of significant long-term \v 
agrees with those for other samples of young stars \citep[e.g.][]{Gag95,
Mar03a, Pi04, Ster95, Mar05}, supporting a scenario in which stars much 
younger than the Sun (i.e. at ages $\la$ 1 Gyr) do not have long-term 
cycles or their cycle amplitudes are much smaller than the solar one.
On the other hand, there is  growing evidence that cycles are present 
in older stars having age comparable to the solar one \citep{He03, Fa04}. 

\begin{figure}
\centerline{\psfig{figure=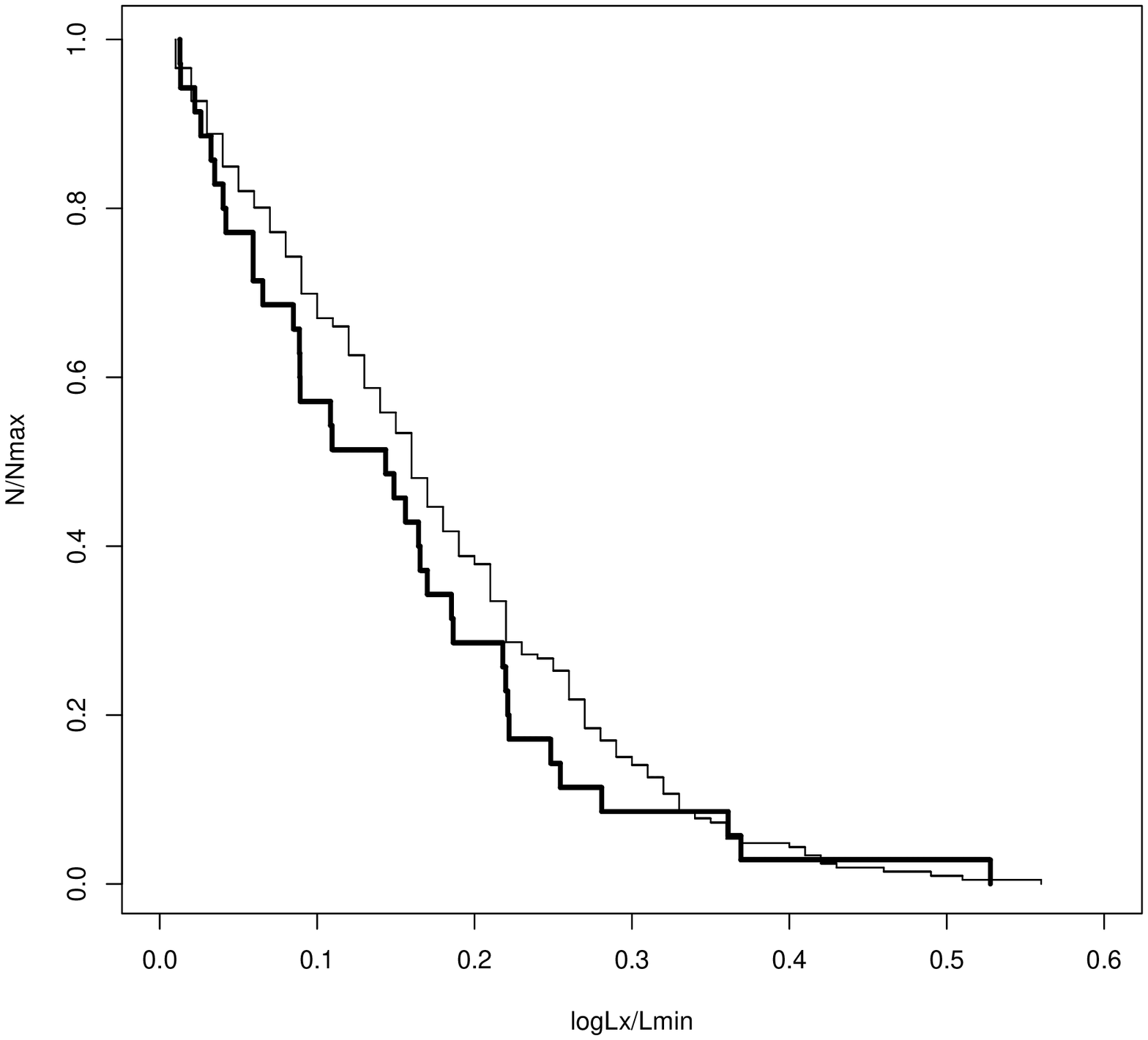,angle=-90,width=9cm}}
\caption{The Time XAD for \N  (thick line) and Pleiades (thin line) 
dF7-dK2 stars.}
\label{GP}
\centerline{\psfig{figure=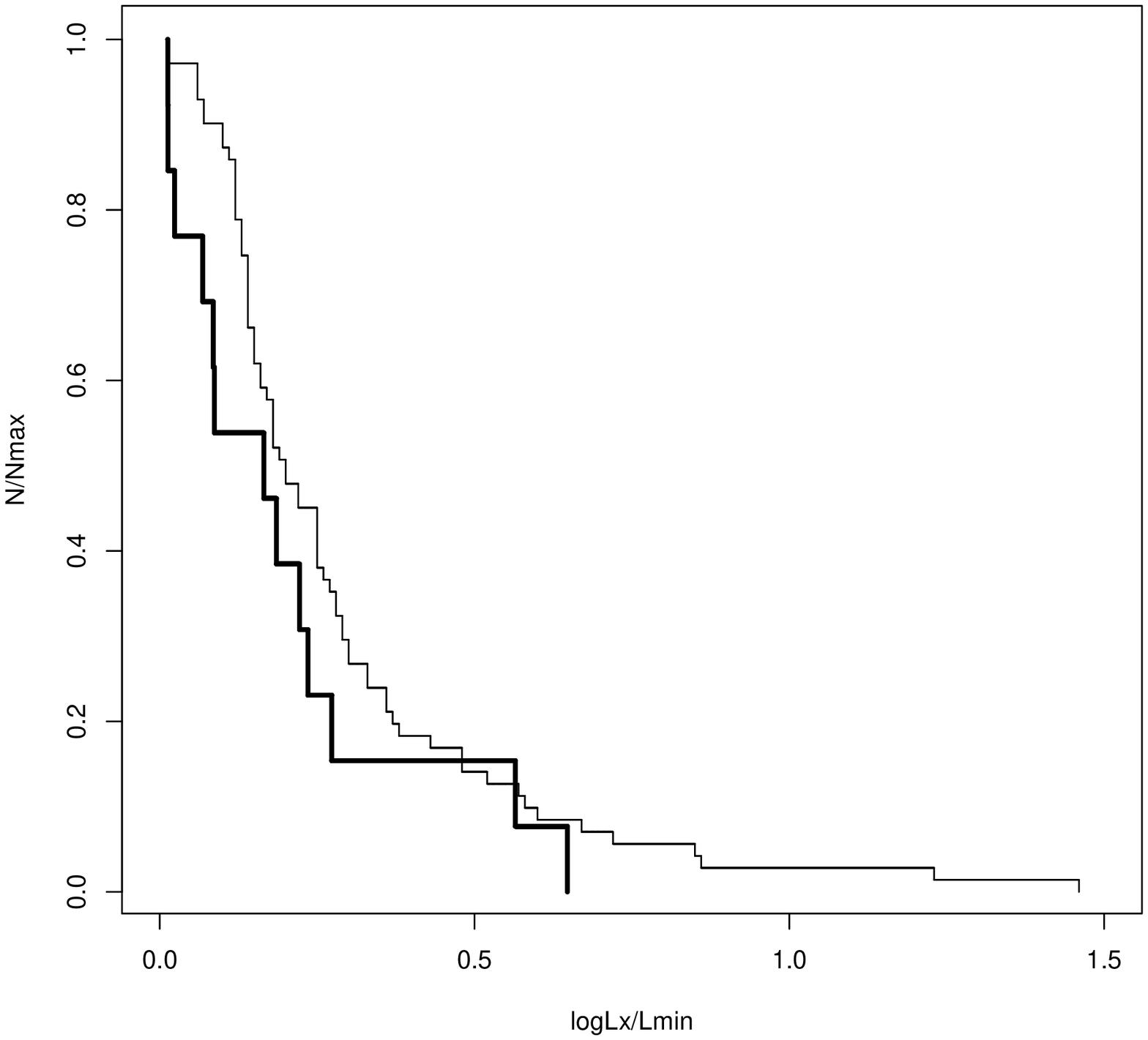,angle=-90,width=9cm}}
\caption{As in Figure \ref{GP} for  dK3-dM stars.}
\label{MP}
\end{figure}

    \subsection{Comparison with the Pleiades}

The X-ray emission level decreases with increasing age as an effect 
of rotational braking \citep[e.g.][]{Mice85, Mice90, Mice96, Ba93, 
Stau94, Ma87}. A decrease by three dex in the median value of L$_x$ 
with a spread of about 1 dex was observed in coeval stars.
How much of this spread is due to \v and both if and how \v properties
change with age is not clear.
Comparing clusters of  a very similar age is very useful for testing the 
evolutionary scenario. 
However ROSAT observations have shown that clusters of the same age 
may have significantly different levels of X-ray emission.
For instance, late type-stars in the Praesepe cluster are much weaker 
emitter in X-rays than the Hyades stars \citep{Ran95}, despite  the 
fact that they have the same age and chemical composition as the Hyades.
Many attempts have been made to explain the observed difference,
but this puzzle has not yet been totally solved. An observation of 
the Praesepe cluster obtained with XMM-Newton \citep{Fra03} supports
the hypothesis that Praesepe may be formed by two merging clusters 
with different ages.

We  compared \N with the slightly younger Pleiades; in particular,  
we compared the Time XADs of dF7-dK2 and dK3-dM members of NGC 2516 
with the analogous ones for the Pleiades \citep{Mar03a}. We converted 
EPIC/pn  count rates to flux  in the 0.1-2.4 keV  ROSAT energy band. 
In deriving X-ray flux we assumed a constant conversion factor of 
4.0 $\times$ 10$^{-12}$ erg cm$^{-2}$/count for the pn camera, 
computed with PIMMS,  for a  1-T Raymond-Smith spectrum  with a 
temperature of logT=6.80 K, and a fixed N$_H$ = 7.5$\times$ 10$^{20}$ cm$^{-2}$
corresponding to the cluster extinction A$_V$ = 0.37 \citep{Jeff01}. 
Figure \ref{GP}  shows the Time XADs of dF7-dK2 in NGC 2516 and Pleiades:
the two distributions appear marginally different, with the Pleiades 
dF7-dK2 amplitude of variations larger than those of \N dF7-dK2. 
However, using the two  sample K-S test, the null hypothesis that
the two distributions came from the same parent distribution, can be
rejected only at a CL $\geq$ 73\%.  
Analogously, Time XADs of dK3-dM in NGC 2516 and Pleiades (Figure \ref{MP}) 
show very similar amplitudes of variations, and the difference is 
marginal since the two distributions differ at a CL $\geq$ 92\%.
These results suggest that the amplitude variations both for dF7-dK2 
and dK3-dM in \N are consistent with the analogous distributions in 
the coeval Pleiades.

       \section{Spectral variability}
Spectral properties  of stellar coronae of the cluster have been 
investigated in \citet{Pil05} by fitting  the X-ray spectra of 
EPIC/pn simultaneously.  
They found that spectra of G, K, and M type stars are  described 
well with one or two thermal components, similar to other clusters, 
such as  Pleiades \citep{Brig03}, Blanco 1 \citep{Pi04}, IC 2391 
\citep{Mar05}, etc.

\begin{figure*}
\centerline{\psfig{figure=4674f7.ps,width=9cm,angle=-90}}
\caption{Spectra of JTH 15510 in the {\em c} (dotted lines), 
{\em d} (dashed lines) and {\em f} (solid lines) Obs.  respectively.}
\label{sp}
\centerline{\psfig{figure=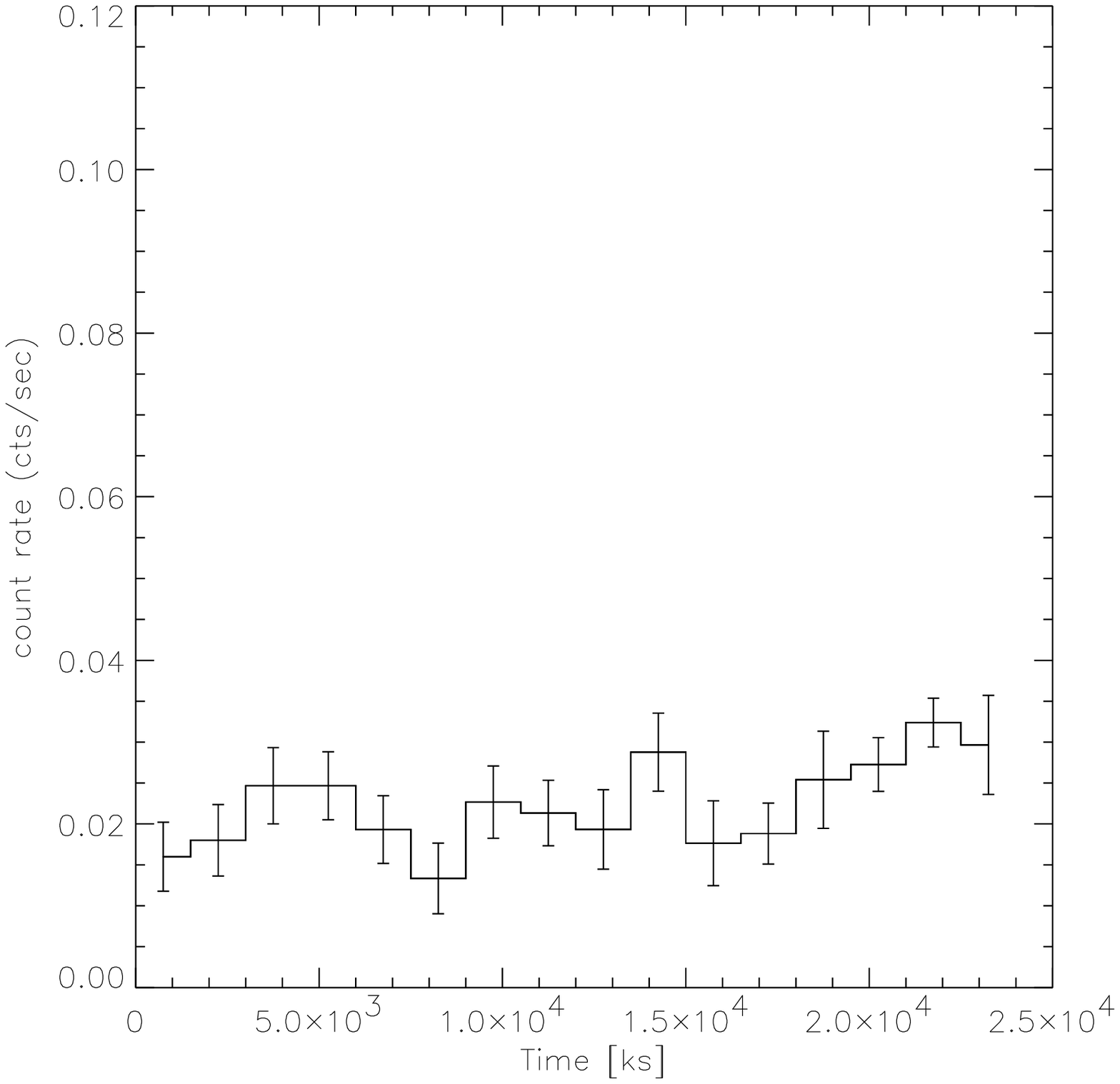,width=5.0cm} \psfig{figure=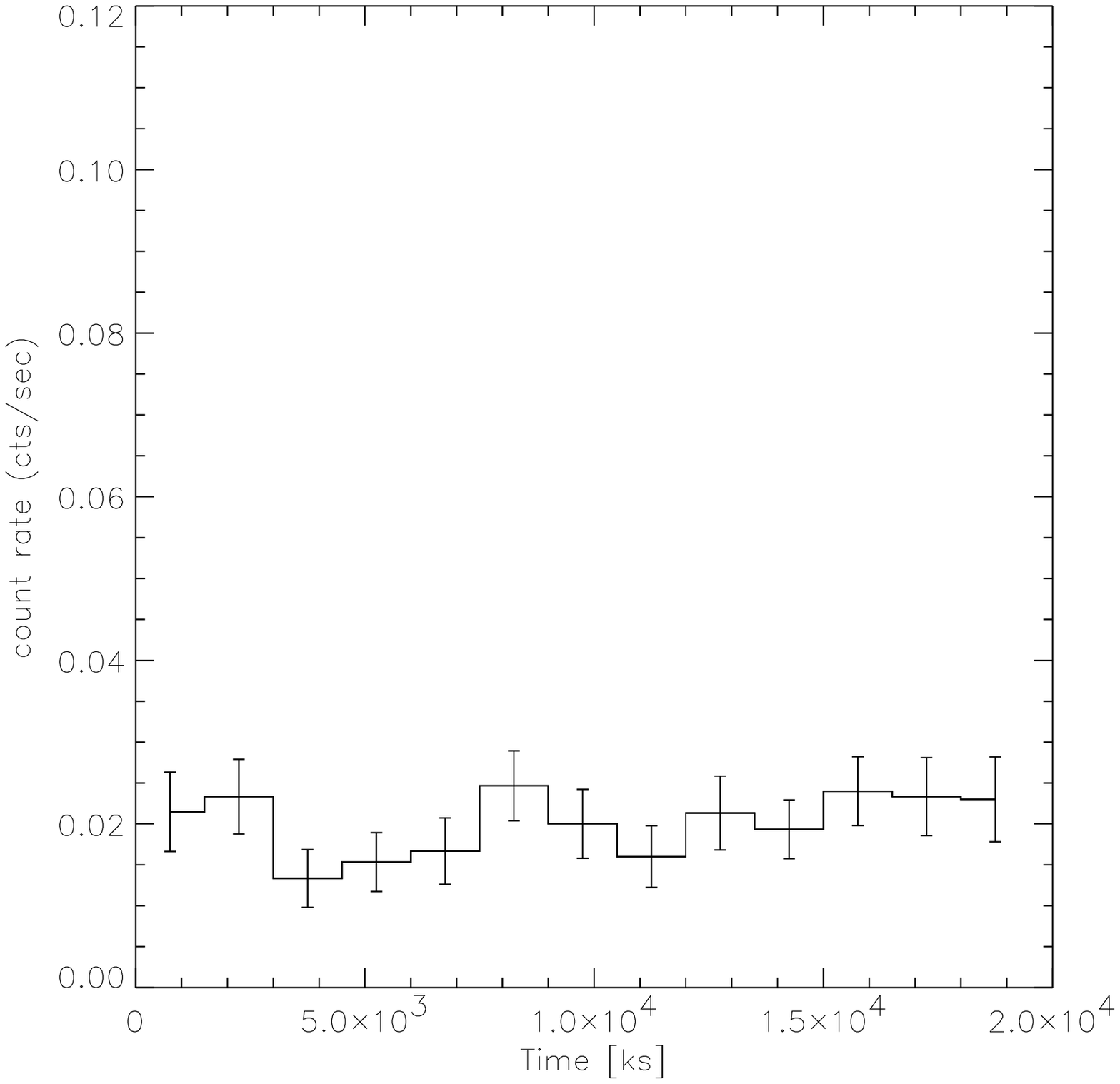,width=5.0cm} \psfig{figure=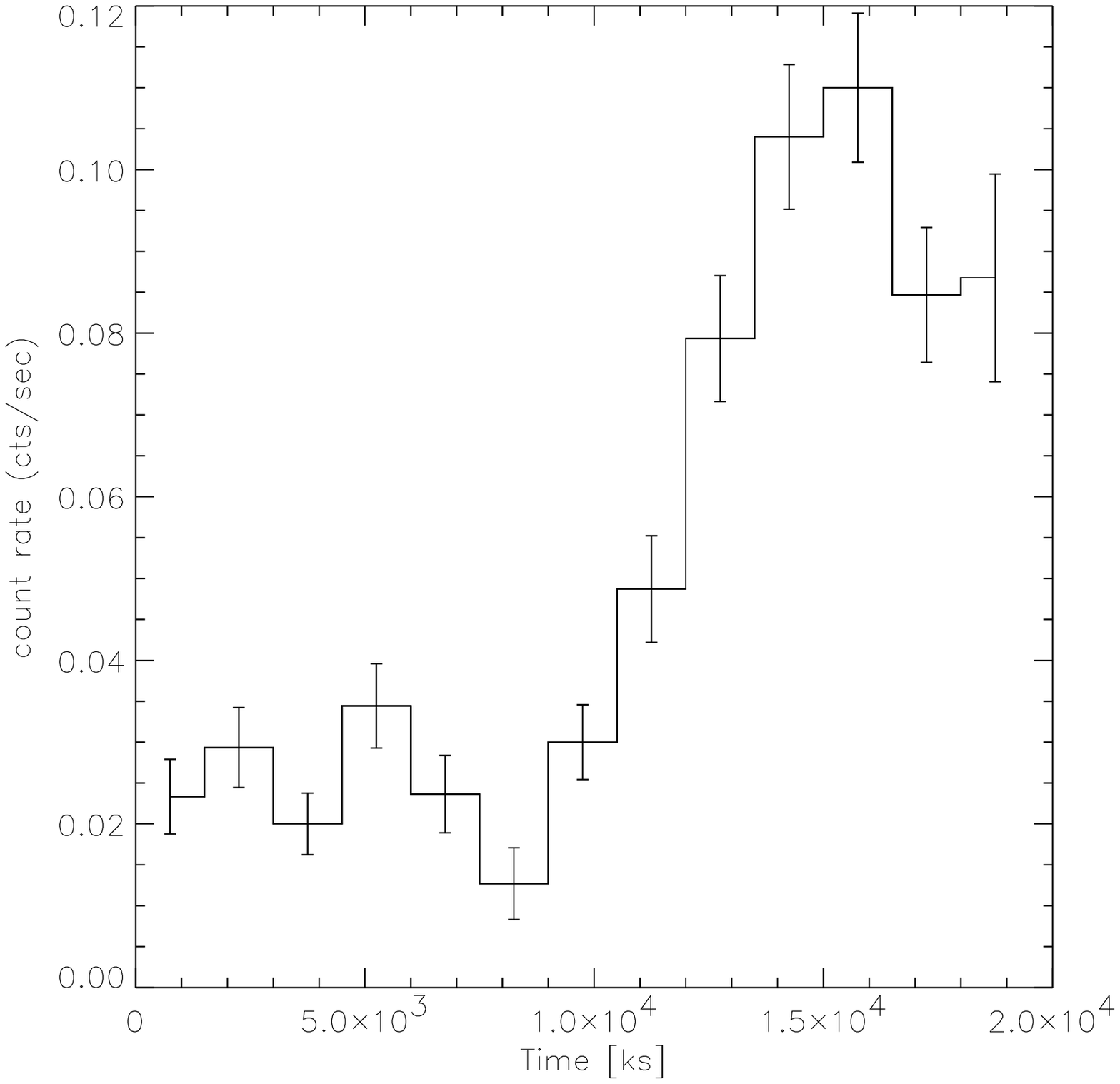,width=5.0cm}}
\caption{From the left: light curves of JTH 15510 in the {\em c}, {\em d} 
and {\em f} Obs. respectively.}
\label{lt}
\end{figure*}

We searched for spectral variations in the cluster members for 
which more than one pn spectra were obtainable.
Due to low counts for most of the sources, this search  was  possible 
only  for three  stars (Table \ref{spv}). 
Two of them  are early-type stars for which stellar structure models 
predict the lack of, or a very thin, convective zone (required to 
generate magnetic activity), and hence the lack of one key ingredient 
for an $\alpha$-$\Omega$ dynamo.
Even though no X-ray emission  is  expected from these stars, the early-type 
JTH 15509 and JTH 15510 are the brightest sources in the EPIC field of view. 
JTH 15510, an A0 star flagged as single  
by \citet{Jef97} in three of the six observations, have sufficient 
counts (more than 350) to obtain three separate spectra (Figure \ref{sp}). 
The two spectra concerning Obs. {\em c} and {\em d} are very similar, and 
simultaneous fits are possible, while the spectrum in the Obs. {\em f} 
is different, implying  different  physical conditions.
As shown in Figure \ref{lt} the spectral \v in  Obs. {\em f}  
corresponds to the presence of a flare in the light-curve, 
while the remaining two  light-curves show  no evidence of 
significant variability.  The  fits  with the 2-T APEC model plus 
a photometric absorption provides for Obs. {\em c} and {\em d}, 
similar temperatures at $\approx$ 0.39 keV and 1.00 keV, with  
EM$_{cold}$/EM$_{hot}$ $\approx$ 1.2 having fixed 
N$_H$ = 6 $\times$ 10$^{20}$ cm$^{-2}$,  and assumed under-solar 
abundances at 0.3 Z$_{\odot}$. To fit  Obs. {\em f}, a third hotter
temperature component  must be added to the model that fits  
Obs. {\em c} and {\em d}. 

The late type star JTH 826 has sufficient counts to obtain 
acceptable spectra in two observations (Obs. {\em a} and {\em f} 
of  Table 1).  Part of  Obs. {\em f} 
a flare that could involve spectral variations. However, this flare 
interested only a small part ($\approx$ 10\%) of the observation, and the 
relatively poor statistic did not allow us to  separately fit flaring 
and quiescent parts of the observation.  The spectra in the two 
observation do not present evidence of spectral variability.

\begin{table*}
\centering
\caption{Stars for which  spectral variability search was possible.}
\bigskip
\begin{tabular}{clcclccc}
\hline
\hline
\multicolumn{1}{c}{JTH} & \multicolumn{1}{c}{RA} &\multicolumn{1}{c}{Dec} & \multicolumn{1}{c}{V}&\multicolumn{1}{c}{B-V}\\
 &  \multicolumn{1}{c}{(J2000)}&\multicolumn{1}{c}{(J2000)} & \\
\hline
  826 &  07:58:43.392 & -60:55:26.76   & 13.84 & 0.818 &  \\
 15510  &  07:58:50.352 & -60:38:38.90 &9.51 & 0.080 \\
 15509  &  07:58:50.616 & -60:49:29.57 &5.80 & -0.090 \\
\hline
\end{tabular}
\label{spv}
\end{table*}

Finally, JTH 15509, a B2 star, is  the optically brightest and hottest
blue star of the cluster; suitable for a search for spectral variability. 
\citet{Dac96}  give for this star an age of the order of one quarter
of that estimated for the cluster ($\sim$ 2.5 $\times$ 10$^{7}$ yr), 
suggesting that it is a blue straggler formed by mass transfer from 
a companion star in a close binary system.
The X-ray emission observed in B stars is generally attributed to 
either intrinsic emission arising from shock heating of the surrounding 
medium by a high-velocity, radiatively driven wind or to the presence 
of an active, late-type companion \citep[e.g.][]{Dac96, Cas94}.
Recent results on early-type stars in the Orion Nebula \citep{Ste05} 
hint that X-ray emission mechanism in hot stars is more complex than 
the simple wind-shock picture, suggesting in particular that magnetic
phenomena may be important even in massive stars.\\
A simultaneous fit  of  four observations (Obs. {\em  a, c, d, f}
in Table 1) with a sum of two APEC models plus a photometric absorption
provides the best fit to the data, as measured with an $\chi^2$ test 
($\chi_{\nu}^2$=1.3 for 82 degrees of freedom).  Assuming under-solar 
abundances at 0.3 and N$_H$ = 6 $\times$ 10$^{20}$ cm$^{-2}$, 
we obtained temperatures of 0.91 keV and 3.29 keV, with a ratio
of the emission measure of EM$_{\rm hot}$/EM$_{\rm cold} \approx$ 7.49.
No evidence of spectral variability is present.

      \section{Summary and conclusions}
We  analysed the \v of members of the open cluster NGC 2516 observed 
with XMM-Newton/EPIC.
Using six XMM-Newton/EPIC observations, we explored X-ray \v on short 
($<$ 1 day), medium (months), and long (years) time scales. 
We  detected 303 distinct detections corresponding to 867 sources, 
and 474 of these sources were identified with 194 members of
the cluster. Stars with spectral type ranging from B-type to M- 
were most detected. 

Variability on short time scales is not very common among \N members 
likely due to the limited statistics of our observations;
the Kolmogorov-Smirnov test applied to all X-ray photon time series of 
detected cluster members shows that only a small fraction (12\%)  of 
the cluster members are variable at a confidence level greater than 99\%, 
suggesting that the X-ray \v does not depend on stellar mass. 

We computed  the time distribution function of the X-ray amplitude
variations for late-type stars in our sample. This distribution 
yields the fraction of time that a star spends with a flux higher, 
by a certain factor, than its minimum value. Our results show that 
the time XADs on the short and medium time scales of solar-like 
(dF7-dK2) stars are very alike. A similar result  has been  found 
for low-mass stars (dK3-dM), suggesting that, on the time scales 
we explored (from $<$ 1 day to 17 months),  the amplitude of \v 
does not change with mass. 
Note that at ages higher than that of this cluster, solar-like 
stars are less variable than low mass stars \citep{Mar02}.
Comparing our data with the ROSAT/PSPC (on time scales of 7-8 yr) 
and ROSAT/HRI (on time scales of 4-5 yr) observations of late-type 
stars of the cluster, we find no evidence for  long-term cyclic \v  
with amplitude similar to the solar one.

The comparison of time XADs for dF7-dK2  and dK3-dM \N stars with 
the corresponding of the coeval Pleiades shows that the amplitude 
variations for the two samples are consistent. 
We  searched for spectral \v in the cluster members for which 
more than one  spectrum was available. We find evidence of spectral 
\v only  in one star due to a  flare in one observation. 

\begin{acknowledgements}{This work is based on observations obtained by XMM-Newton, 
an ESA science mission with instruments and contributions directly funded by ESA 
Member States and the USA (NASA). We acknowledge financial support from MIUR and the 
anonymous referee.} \end{acknowledgements}
\bibliographystyle{aa}
\bibliography{4674}
\end{document}